\documentclass[aps,prl,superscriptaddress,twocolumn,showpacs,10pt]{revtex4-2}
\usepackage [utf8]{inputenc} 
\usepackage{graphicx, subfigure, amsmath, overpic, color}
\usepackage[normalem]{ulem}
\usepackage[section]{placeins}

\newcommand{\ieap}{Institut f\"ur Experimentelle und Angewandte Physik, Christian-Albrechts-Universit\"at zu Kiel, D-24098 Kiel, Germany}
\newcommand{\dtu}{Department of Physics, Technical University of Denmark, DK-2800 Kongens Lyngby, Denmark} 
\newcommand{\cemes}{CEMES, Universit\'e de Toulouse, CNRS, 29 rue Jeanne Marvig, F-31055 Toulouse, France}

\newcommand{\angstrom}{\textup{\AA}}

\begin{document}

\title{Current shot noise in atomic contacts: Fe and FeH$_2$ between Au electrodes}
 
\author{Michael Mohr} \affiliation{\ieap}
\author{Alexander Weismann} \affiliation{\ieap}
\author{Dongzhe Li} \affiliation{\dtu}\affiliation{\cemes}
\author{Mads Brandbyge} \affiliation{\dtu} 
\author{Richard Berndt} \affiliation{\ieap}
 
\begin{abstract}
Single Fe atoms on Au(111) surfaces were hydrogenated and dehydrogenated with the Au tip of a low-temperature scanning tunneling microscope (STM). Fe and FeH$_2$ were contacted with the tip of the microscope and show distinctly different evolutions of the conductance with the tip-substrate distance. The current shot noise of these contacts has been measured and indicates a single relevant conductance channel with the spin-polarized transmission. For FeH$_2$ the spin polarization reaches values up to 80\% for low conductances and is reduced if the tip-surface distance is decreased. These observations are partially reproduced using density functional theory (DFT) based transport calculations. We suggest that the quantum motion of the hydrogen atoms, which is not taken into account in our DFT modeling, may have a significant effect on the results.

\end{abstract}
 
 
\maketitle
 \section{I. Introduction}
The charge and spin transport through atomic and mesoscopic nanostructures is subject to genuine quantum effects.
For example, electron correlations due to the Pauli principle --- becoming increasingly important for reduced dimensions --- impinge on current fluctuations \cite{blanter}.
As a result, the quantum shot noise of the electronic current through nanoscale objects can be significantly reduced compared to the Poissonian shot noise that occurs for uncorrelated charge carriers.
Vice versa, noise measurements may be used to characterize the spin-polarization of the current and the quantum states that are involved in transport.
 
The possibility of highly spin-polarized transport through atomic contacts between ferromagnetic metals has initially been investigated using conductance measurements.
However, contradictory results were obtained, possibly due to the involvement of contaminants \cite{untie04}.
More recently, break junction experiments have been extended to measurements of the shot noise.
This noise  has been intensely studied for homogeneous junctions from various materials, including noble metals \cite{VanDenBrom_PRL, chen_enhanced_2014, wheeler2010shot, Vardimon, RBG_Kumar_PRL, MKu} and ferromagnets \cite{vardimon2016orbital} and for molecules like benzene \citep{Ruitenbeek_Benzene}, D$_2$ \cite{RBG_Djukic_single_molecule}, 1, 4-benzenedithiol \cite{karimi2016shot}, and vanadocene \cite{pal2018electronic} placed between electrodes.
In addition, a symmetry-based mechanism to fully block the majority-spin conductance by joining two ferromagnetic electrodes via a $\pi$-conjugated molecule has recently been proposed from \textit{ab initio} theory \cite{smogunov2015}.
 
Fe, Co, and Ni display maxima in conductance histograms at conductances $G > \text{G}_0=2e^2/h$ being the quantum of conductance \cite{untie04, Smit2002, vardimon2016orbital}.
This indicates that electron transport involves several partially open conductance channels.
Moreover, the absence of a maximum at ${1 \over 2} $G$_0$ did not suggest fully spin-polarized transport \cite{untie04}.
Gas molecules added to the junctions modified the transport properties.
For nickel electrodes, the introduction of small gas molecules was reported to drastically increase the measured spin polarization \cite{vardimon2015indication, Dongzhe-N2-2019}.
This effect has been attributed to orbital symmetries and a resulting reduction of the number of transport channels. 
Hydrogen and deuterium have been studied experimentally and theoretically in a number of cases, sometimes along with vibrational spectroscopy and noise measurements \cite{Smit2002, Garcia2004, Djukic2006, Thijssen2006, kig07, Kristensen2009, Kiguchi2010, Li2015, Hauho15}.

Investigations of shot noise with STM are still fairly few in number \cite{Schoenenberger, nonoise, chen12, abu, peters17, Mohr_Nioc, Mohr_TOTA, bastiaans_charge_2018}.
Here we present results from  single Fe atoms on Au(111) surfaces that are hydrogenated in a low-temperature scanning tunneling microscope (STM). 
The presence of hydrogen modifies the evolutions of the conductance with the tip-substrate distance and also affects the shot noise of the current. 
We find fairly large spin polarizations of up to 80\% at low conductances for FeH$_2$ and lower values at lager conductances and also for pristine Fe.
Transport calculations based on density functional theory (DFT) reproduce important aspects of the experimental data.
However, no quantitative agreement is obtained and we suggest that the quantum motion of the hydrogen atoms may be at the origin of the deviation.
 
\section{II. Experimental details}
 
We used a low-temperature STM operated at 4.4~K in ultra-high vacuum (UHV) with a base pressure below $5 \times 10^{-11}$~mbar.
Au(111) single crystal surfaces were prepared by cycles of Ar ion sputtering and annealing. 
Tips were etched from W wire and prepared in UHV by annealing. 
After mounting into the STM they were indented into the Au crystal to coat their apex with gold.
Finally the tip was gently brought into contact with the surface until single Au atoms were deposited from the tip and stable contacts with conductances of $G\approx 1\,\text{G}_0$ were achieved.
Single Fe atoms were deposited onto the cold Au sample by evaporation from an Fe covered W filament.
Contacts to single atoms were achieved by centering the STM tip above them, switching off the current feedback, and bringing the tip closer until a rapid rise was observed in current-displacement curves $I(\Delta z),$ which were recorded simultaneously.
Current noise was measured with a setup described in Ref.~\cite{abu}. 
For this, the contact was biased using a battery-driven low-noise current source.
The voltage noise of the contact was measured using two amplifiers in parallel and cross-correlated to reduce amplifier noise.
The resulting spectrum of the noise power density was averaged at frequencies, where white noise was present (usually between 80~kHz and 100 ~kHz).
 
To characterize a single contact, noise measurements were performed for several different currents.
In addition to the noise, the DC voltage drop over the contact was recorded to calculate its conductance.
For further analysis, we only used data from contacts whose conductances were identical on the increasing and decreasing part of the bias current ramp. 
For comparison, we determined the conductance from thermal noise $S_\theta = 4 k_B T G$ ($T:$ temperature, $k_B:$ Boltzmann constant) measured at $I=0$.
These conductances agreed within $\pm 5$~\%.
 
\section{III. Experimental results}
 
\subsection{A. STM images of Fe atoms and Fe hydrides}
 
\begin{figure}[hbt]	
		\includegraphics[width=0.8\linewidth]{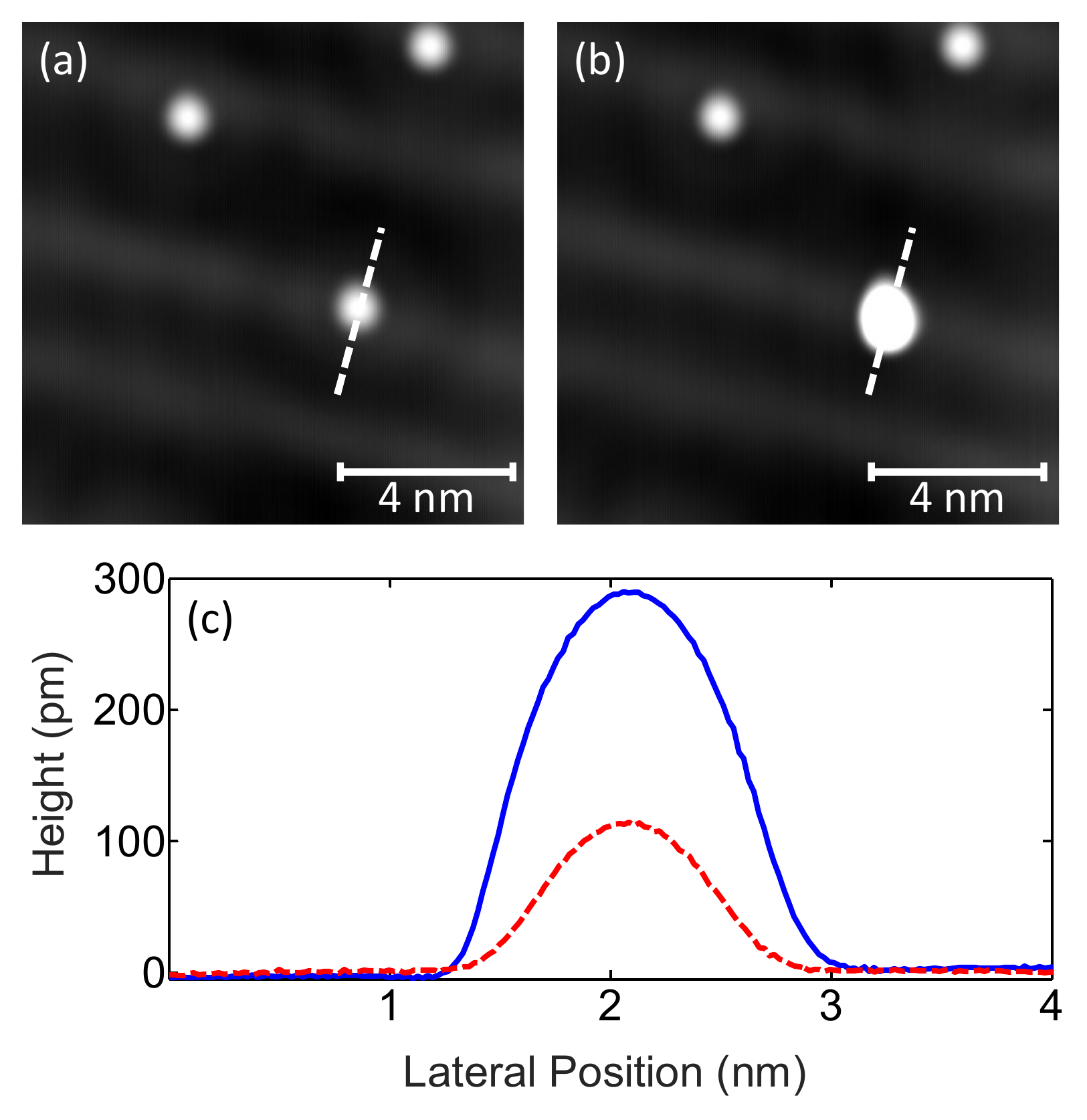}
		\caption{
		(a) Topograph (14$\times$14~nm$^2$, 129~mV, 50~pA) showing three Fe adatoms on Au(111).
		(b) Image of the same area recorded after contacting the atom near the center with the STM tip.
		(c) Cross-sectional profiles of the pristine (dashed) and hydrogenated Fe adatom (solid).
		The apparent height increases from $\approx 110$ to $\approx 280$~pm.}
\label{topo}
\end{figure}

After the evaporation of a submonolayer amount of Fe, two kinds of isolated protrusions were observed on the Au(111) substrate, which have apparent heights of $\approx$~110~pm [Fig.~\ref{topo}(a)] and 150~pm (not shown).
The ratio of their abundances was approximately 10:1.
When contacted with the STM tip the less abundant structures turned out to be unstable and their apparent height consistently increased to 280~pm.
In contrast, the 110~pm structures often remained unchanged when contacted.
Occasionally, however, they were also converted to 280~pm signatures.
An example of this conversion is shown in Fig.~\ref{topo}.
After many (on the order of 10) 110~pm to 280~pm conversions with the same tip, no more changes took place and the contacted 110~pm protrusions remained stable.
 
We interpret the observed structures as Fe adatoms and Fe-Hydrogen complexes.
During contact formation, hydrogen can be transferred from the tip to the lower protrusions which increases their apparent height.
After a number of repeated transfer processes, hydrogen is depleted from the tip apex.
However, hydrogen transfer from the tip was observed again after an extended waiting time (several hours), suggesting that hydrogen from the tip shaft had moved to the tip apex. 
 
We thus attribute the 110~pm structures to Fe adatoms due to their high abundance right after sample preparation and their low apparent height.
According to our DFT calculations discussed below, complexes with hydrogen appear taller.
Furthermore, we found that sample voltage pulses of 1.3~V applied after disabling the feedback loop at 129~mV and 50~nA reliably changed the higher structures back to an apparent height of 110~pm.
 
The 280~pm structures are attributed to Fe with a H$_2$ molecule that is oriented perpendicular to the substrate.
This interpretation is suggested by DFT results (\textit{vide infra}). 
It is also consistent with the observed depletion and replenishing of the tip apex by repeated transfer and waiting.
 
Related observations have been reported for the transition metal adatoms Ce \cite{piv07}, La and Cr \cite{ternes2008spectroscopic}, Ti \cite{nat13}, Co \cite{serrate2014enhanced, dubout2015controlling, jacobson2017potential}, W \cite{sal14}, and Fe \cite{kha15} where hydrogen was found to affect the apparent height and other properties like the Kondo effect. 
In contrast, metals M from group III of the periodic table, which form MH$_3$ compounds, Fe prefers to bind one or two H atoms \cite{hydride}.
 
Finally, we observed that the FeH$_2$ complexes changed to 150~pm on a timescale of hours.
At a slightly higher sample temperature of $\approx 20$~K, this change occurred within minutes. 
This may be interpreted as a conversion from FeH$_2$, where H$_2$ is perpendicular on the substrate, to a parallel orientation or to FeH\@. We did not further investigate this aspect because the 150~pm structures were unstable when contacted. 

\subsection{B. Conductance measurements}
 
\begin{figure}[hbt]	
	\includegraphics[width=0.8\linewidth]{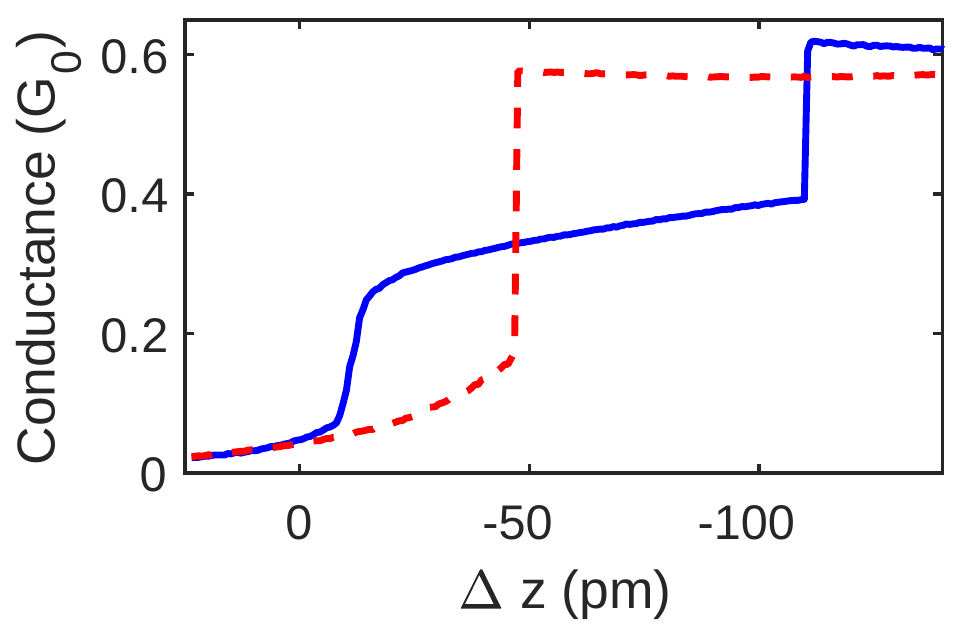}
	\caption{Conductance versus tip displacement $\Delta z$ recorded from a Fe adatom (dashed line) and FeH$_2$ (solid line) at $V=129$~mV.
$\Delta z=0$ corresponds to a current $I=500$~nA\@.}
\label{iz}
\end{figure}

Figure~\ref{iz} shows the evolution of conductances $G$ as a function of the vertical tip displacement $\Delta z$ for contacts to Fe (dashed) and FeH$_2$ (solid).
Following an exponential increase at larger distances, the data from Fe exhibit a sharp jump to contact ($G=0.58 \,\text{G}_0$) and a plateau of nearly constant $G$.
In contrast, the conductance of FeH$_2$ varies more smoothly at larger tip-adatom separations with a rise to $G\approx 0.25 \,\text{G}_0$ at $\Delta z \approx -14$~pm and a further sub-exponential increase.
At further reduction of tip-sample distance ($\Delta z \approx -110$~pm), an additional steep rise of the conductance could be observed in some cases.
The abruptness of the rise and the conductance $(\approx 0.6\, \text{G}_0)$ of the fairly flat plateau for large $\Delta z$ are very similar to the observations from pristine Fe contacts.
It turned out that contacts to FeH$_2$ are stable and exhibit no hysteresis as long as the abrupt jump to the second conductance plateau does not occur.
When the jump has taken place, however, the conductance-distance curves measured during tip approach and retraction are usually no longer identical and show hysteretic behavior. 
The distance between jumps during approach and retraction varies between 0 and 120~pm. 
It should be noted, that in some cases no second plateau was observed for FeH$_2$.
 
Repeated measurements on different pristine Fe atoms with different tips showed a scatter of the plateau conductances between 0.4 and 0.75~G$_0$.
For FeH$_2$ the conductance of the second plateau varied between 0.37 and 0.89~G$_0$.
We hint that this scatter reflects the unknown atomic structure of the tip apex and details of the Fe adsorption site on the Au(111) substrate.
 
\subsection{C. Noise measurements}

\begin{figure}[hbt]
	\centering
	\includegraphics[width=0.8\columnwidth]{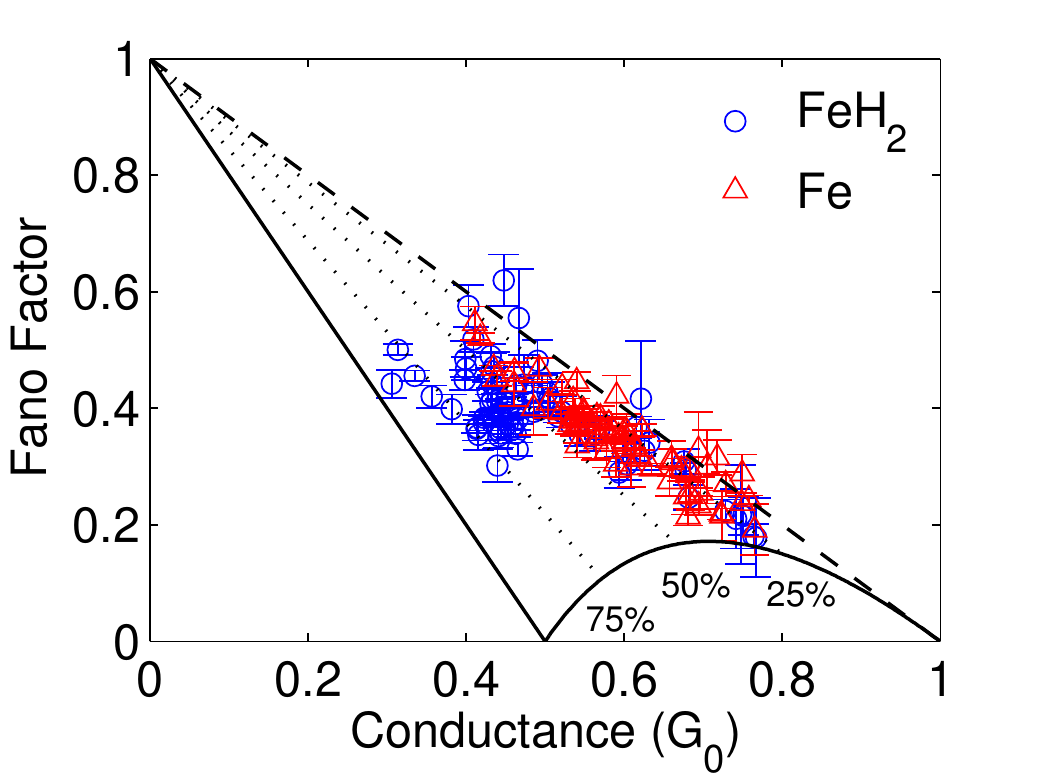}
	\caption{
	Fano factors versus conductances of contacts to Fe (red triangles) and FeH$_2$ (blue circles) on Au(111) determined from the noise measured at $T=4.5$~K with different tips.
	Bars indicate the 95\% confidence interval of $F$.
	Solid and dashed lines indicate the minimal Fano factor for spin-polarized and spin-degenerate conduction channels, respectively.
	Dotted lines indicate lower boundaries of spin polarization.
	Nearly all Fe data sets suggest spin polarized transport with a maximal polarization of 50~\%.
	FeH$_2$ contacts exhibit higher degrees of polarization often exceeding 60~\% for conductances below 0.5~G$_0$.}
	\label{alle}
\end{figure}

Spectra of the current noise $S$ were measured on Fe atoms and FeH$_2$ molecules.
The excess noise $\Delta S$ was obtained as \cite{Lesovik, abu}
\begin{equation}
\Delta S = S - S_\theta = F\left[ S_0 \, \coth\left( \frac{S_0}{S_{\theta}}\right)-S_{\theta} \right],\label{leso}
\end{equation}
where $S_\theta=4k_BTG$ is the thermal noise, $S_0 = 2 e I$ is the Poisson (full) shot noise and $F$ is the Fano factor.
$F$ describes the noise reduction of quantum systems due to the anti-correlations of transmission events.
Equation \ref{leso} was used to fit the measured noise $S(I)$ treating $F$ and $T$ as adjustable parameters.
The average fit result of $T$ from all data presented in this paper was $4.2$~K (standard deviation $1.2$~K).
Figure \ref{alle} shows the obtained Fano factors of Fe (red triangles) and FeH$_2$ (blue circles).
While there is substantial scatter, the Fe data appear to obey a linear relationship between $F$ and $G$.
For FeH$_2$ Fano factors tend to be smaller than those of Fe contacts for similar $G$.
The difference between the Fano factors of FeH$_2$ and Fe contacts is most obvious for low conductances.

While Fe atoms may in principle provide several conductance channels the Au tip is expected to act as an orbital filter \cite{abu}.  
Indeed, the conductance of single-atom Au contacts is predominantly due to a single $s$-like state \cite{sir96}.
Transport calculations predict that the Au tip used in the present experiments similarly provides a single transport channel \cite{jac6, roc7, abu}. 
This interpretation also matches the results of break junction experiments with Fe contacts \cite{vardimon2016orbital}.  

Next, we consider subchannels for spin-up and spin-down electrons with the transmission probabilities $\tau_\uparrow$ and $\tau_\downarrow$.
The spin polarization of the transmission is defined as
\begin{equation}
P=\frac{\tau_\uparrow-\tau_\downarrow}{\tau_\uparrow+\tau_\downarrow}
\end{equation}
and the Fano factor is 
\begin{equation}
F = \frac{ \sum_i \tau_i( 1 - \tau_i ) }{ \sum_i \tau_i },
\end{equation}
where $i=\uparrow, \downarrow$ are the spin directions.

Solid and dashed lines in Fig.~\ref{alle} indicate $F$ for fully spin-polarized $(P = 1)$ and unpolarized $(P = 0)$ transmission.
Dotted lines show a few intermediate values of $P$.
A vast majority of data lie below the dashed line showing that there is a degree of spin polarization.
The upper limit of $P$ is close to 50~\%, similar to previous results \cite{abu}. 
The presence of hydrogen has a limited effect on $P$ at large conductances.
At lower conductances $G<0.5~$G$_0$, however, hydrogen increases the spin polarization reaching values up to 90~\%.

\begin{figure}[hbt]
	\centering
		\includegraphics[width=0.8\columnwidth]{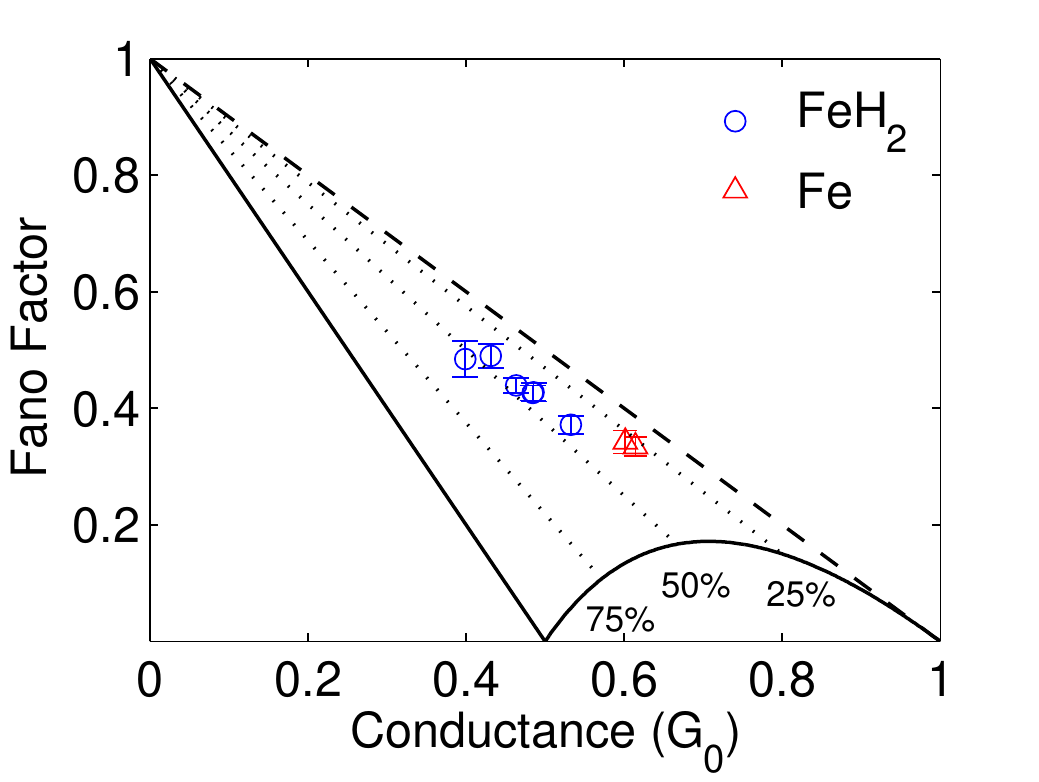}
	\caption{
	Sequence of Fano factors observed from the Fe atom of Fig.~\ref{topo} that undergoes hydrogenation to FeH$_2.$
	Two measurements on the pristine atom indicate $P \sim$ 30\%.
	After $\approx 1$~h of repeated measurements, the atom was hydrogenated during contact formation.
	The resulting FeH$_2$ molecule displays a significantly higher polarization between $\approx 40$ and  60~\%.}
	\label{same_atom}
\end{figure}

Since Figure~\ref{alle} shows results from different experimental runs with presumably widely different tip apices, the scatter obscures the effect of H$_2$ on $P$~to some extent. 
It is more readily apparent in noise data from an Fe atom before (red triangles) and after hydrogenation (blue circles), measured with the same tip (Fig.~\ref{same_atom}).
The Fe contact initially exhibits a spin polarization of 30~\% and repeated measurements lead to very similar results.
One hour later - which apparently resulted in the presence of hydrogen at the tip apex - another contact process changed the apparent height to 280~pm indicating hydrogenation. 
A subsequent contact showed a reduced conductance and an increased spin polarization which was confirmed by further measurements at slightly different conductance values.
 
\begin{figure}
	\centering
		\includegraphics[width=0.8\columnwidth]{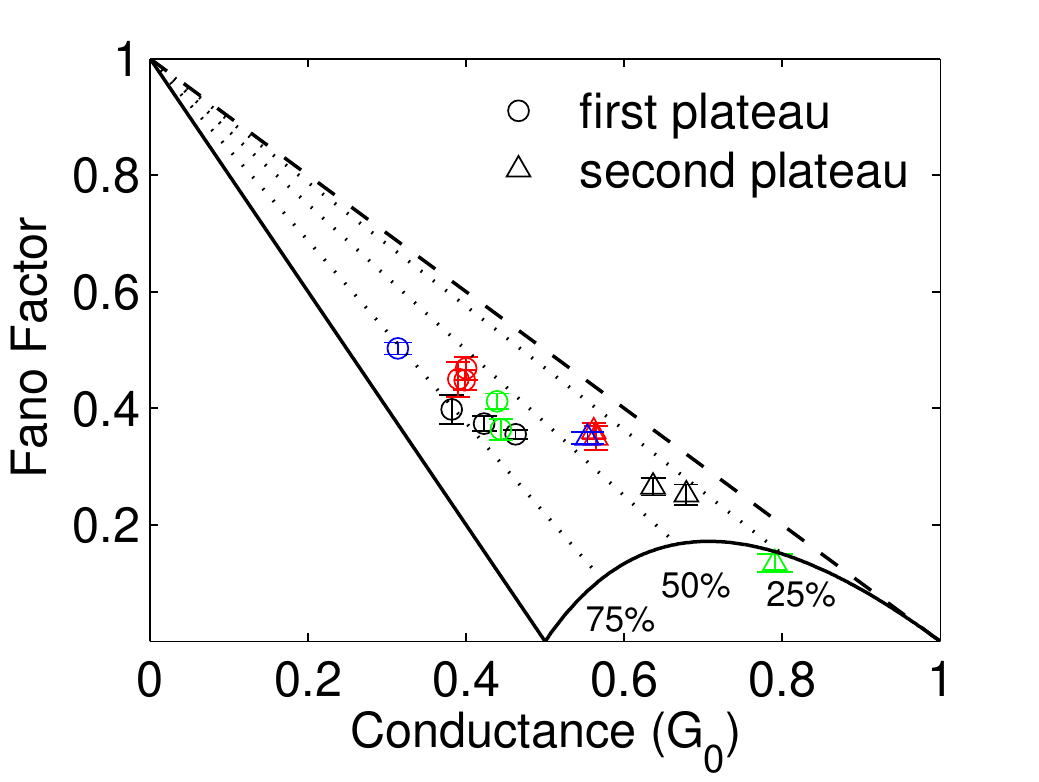}
	\caption{
	Four independent data sets of Fano factors that were measured on the first and the second conductance plateau of FeH$_2$.
	Each color corresponds to one dataset measured on the same atom with the same tip.
	Circles (triangles) indicate values measured on the first (second) plateau.}
	\label{noise2}
\end{figure}

Finally, we investigated contacts that could be stably operated on both conductance plateaus.
The Fano factors at $\Delta z$ corresponding to the first (circles) and second (triangles) plateau are shown in Fig.~\ref{noise2}.
While all data display a significant spin polarization, $P$ is substantially higher on the first plateau at low conductance.
The polarization on the high-conductance plateau is similar to the results from pristine Fe. 

A simple model that would be consistent with these observations is that H$_2$ is initially located close to the axis of the contact but is expelled from this position by the Au tip, which then directly contacts Fe.
However, the transport calculations below favor a different mechanism.

\section{IV. Theoretical modeling}

To help interpret the experiments and establish a connection between the transport features and the atomic geometry, we employed first-principles modeling of the contacts and during the formation process. 
We used DFT with the PBE exchange correlation functional  \cite{perdew1996generalized} including a van der Waals correctio \cite{grimme2006} in the usual Born-Oppenheimer approximation with frozen atomic configurations, as implemented in the $\textsc{Siesta}$ \cite{soler2002siesta,garcia2020siesta} code.
Pseudopotentials were obtained from Ref.~\cite{rivero2015} and the mesh cutoff (filter cutoff) values were set to 400 Ry (300 Ry). 
A 4$\times$4 (111) surface unit cell was adopted along with a four-atom tip-pyramid mounted on the opposite (111) surface, and the 2D Brillouin zone was sampled with 4$\times$4 $k$-points mesh.
To simulate the tip approach, we decreased the tip-sample distance in a step-wise manner ($\Delta z$), and optimized the junction geometry at every step. 
In the atomic relaxation, the Fe, H, Au tip, and outermost substrate layers were relaxed ($<$ 0.01~eV/\angstrom). 
For every step subsequent spin transport calculations were performed using the $\textsc{Transiesta}$ non-equilibrium Green's function method (DFT+NEGF)~\cite{brandbyge2002density,papior2017improvements}.
Transmissions and Fano factors were obtained from the eigenchannel transmissions [$\tau_k(n)$] sampled on a 30$\times$30 $k$-point mesh, using $\langle \tau_k(n)\rangle_k$ and $\langle \tau_k(n)(1-\tau_k(n))\rangle_k$ and the $\textsc{SISL}$ postprocessing tool \cite{zerothi_sisl}.


\begin{figure}
 \centering
 \includegraphics[width=0.9\columnwidth]{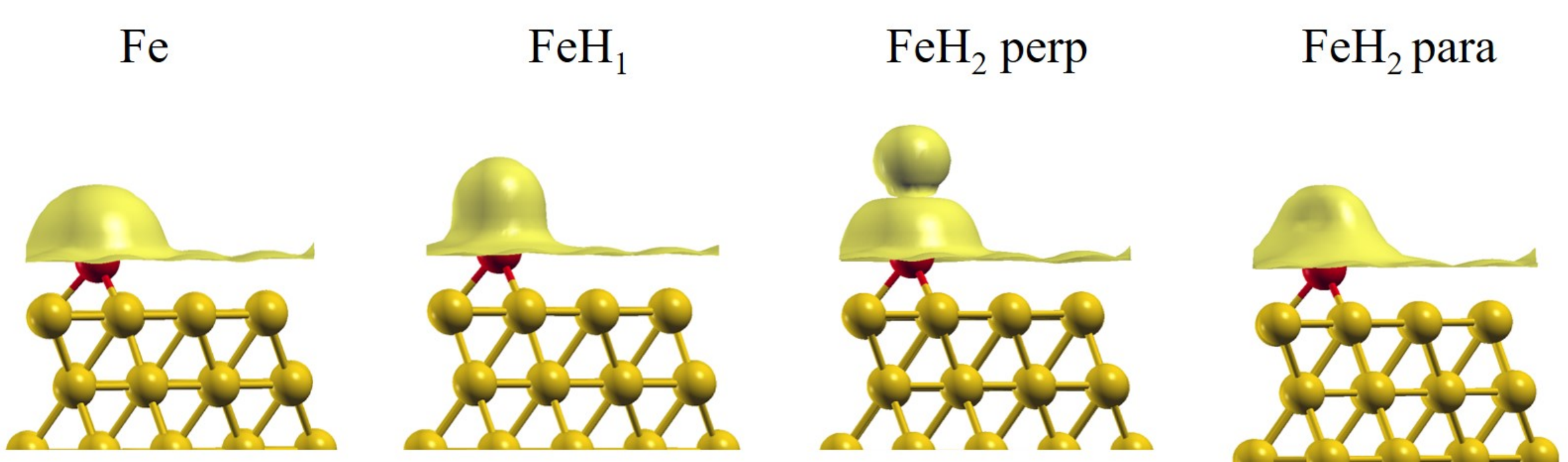}
 \caption{Contour surface of the local density of states at the Fermi energy for Fe, FeH$_1$, FeH$_2$ perpendicular, and FeH$_2$ parallel configurations.}
 \label{theory_fig1}
\end{figure}

\section{V. Theoretical results}

In the experimental STM topographs, the hydrogenated Fe adatoms appear significantly higher than pristine ones (Fig.~\ref{topo}). In Fig.~\ref{theory_fig1} we show the isosurface of the calculated local density of states (LDOS) around the Fermi energy for Fe, FeH$_1$, and the perpendicular and parallel configurations of FeH$_2$ in the absence of a STM tip. 
The perpendicular configuration provides the best match with the experimental result from the hydrogenated Fe adatoms. We find that the FeH$_2$ parallel is energetically more favorable than the perpendicular one with an energy difference of about $\Delta E = 0.95$ eV when the tip is not present. However, importantly, we note that when the interaction with the STM tip is introduced, this $\Delta E$ is significantly reduced to $\Delta E \approx$ 0.18~eV.
 
We considered six initial, stable configurations of the junction prior to contact formation:
Clean Fe, FeH$_1$, and two different FeH$_2$ contacts corresponding to broken (perp I) and intact (perp II) H--H bonds, respectively (Fig.~\ref{theory_fig2}).
We further considered two parallel configurations (see Fig.~\ref{theory_sup3} in Appendix A). The corresponding conductances versus tip displacements are shown in Fig.~\ref{theory_fig2}(a).

\begin{figure}
\centering
\includegraphics[width=0.8\columnwidth]{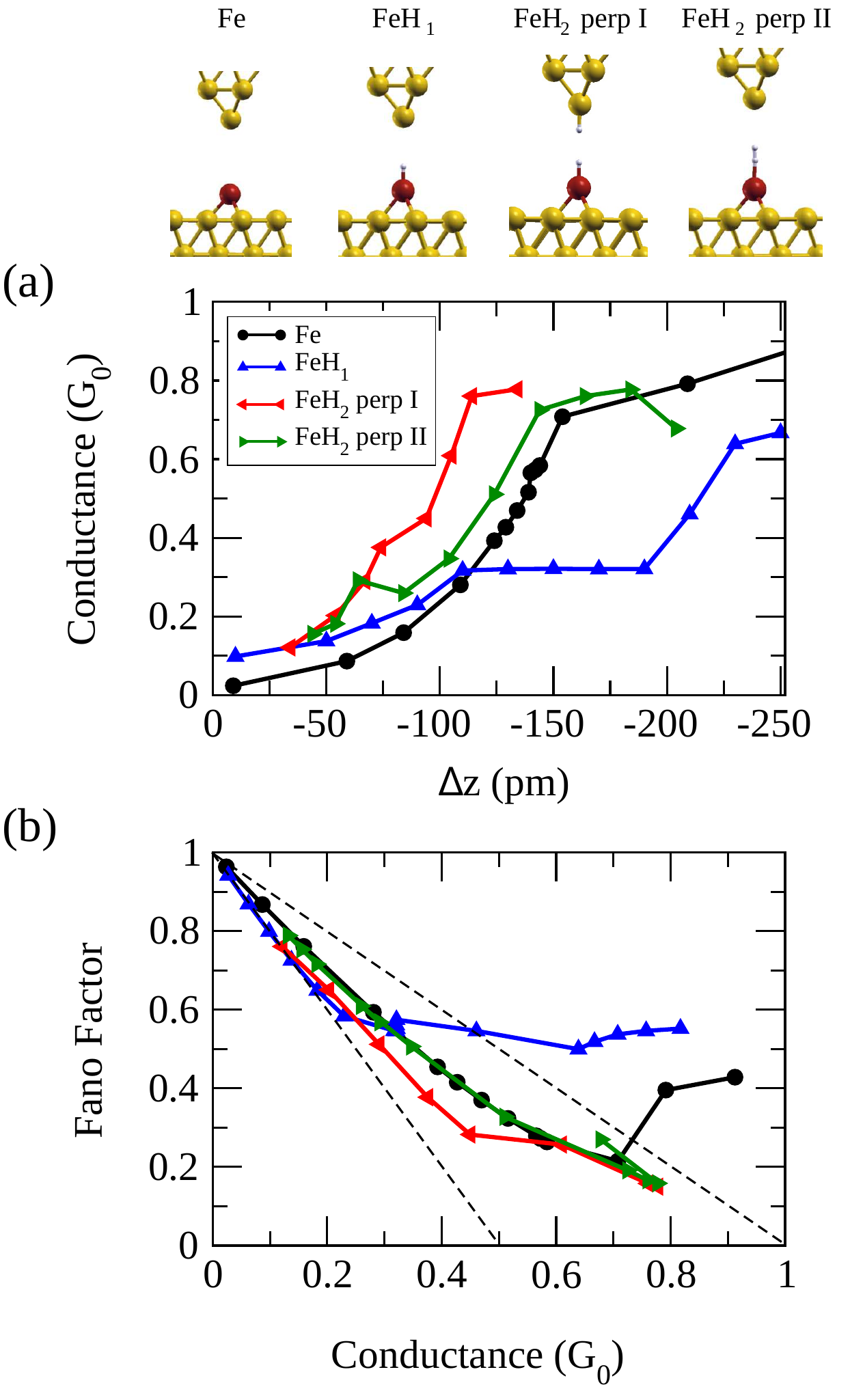}
\caption{Side views of calculated initial positions ($\Delta z=0$) of Fe (red spheres), hydrogen (white spheres), and Au (yellow spheres) for four different types of atomic contacts are shown on top of the figure. Note that the $\Delta z=0$ (reference point) corresponds to the tunneling conductance of Fe contact smaller than 0.01G$_0$. Conductance versus tip displacement $\Delta z$ (a) and Fano factors versus conductances (b) of atomic contacts to Fe (black circles), FeH$_1$ (blue up triangles), FeH$_2$ perpendicular I (red left triangles) and FeH$_2$ perpendicular II (green right triangles) on Au(111).}
\label{theory_fig2}
\end{figure}
 
For clean Fe, we find a conductance plateau about 0.7~G$_0$, as shown in Fig.~\ref{theory_fig2}(a), slightly higher than found experimentally. For both FeH$_1$ and FeH$_2$, two plateaus occur, in rough agreement with the experimental measurements. From the shot noise calculations, we extract a transport-spin polarization, $P$, of the clean Fe contact of about 40\%, cf.\ Fig.~\ref{theory_fig2}(b). 
 
To achieve the perp I configuration we have to start with hydrogen atoms placed at the tip and Fe, respectively. We find that the perp I configuration has a slightly lower energy than perp II (Fig.~\ref{theory_sup5} in Appendix A). We include the broken-bond perp I configuration since the energy of the perpendicular configurations will converge during the tip approach. To this end we note that nuclear quantum effects (NQEs) due to the small mass of hydrogen are absent in our calculations. The NQE of the hydrogen could involve tunneling and fluctuations between these configurations \cite{Lauhon2000, Kiguchi2010, Fang2019}. 

The conductance versus tip displacement data display a short ``plateau" around 0.3 and 0.4~G$_0$ for perp II and perp I, respectively. This conductance value is quite close to the experimentally observed conductance for the plateau associated with FeH$_2$. On the other hand, a long, clear, plateau around this value is also seen in the calculations for the FeH$_1$ contact. 
It is the yielding at contact and subsequent relaxation of the Fe-substrate bonds which leads to the longer plateau for FeH (Fig.~\ref{theory_sup11} in Appendix A). \@
It can not be ruled out that that this region of conductance in reality involve fluctuations between all three configurations, where the hydrogen on the tip is fluctuating in and out of the contact region yielding similar conductances as the experimental plateau. The fluctuations between two levels induced by the current have been observed in H$_2$-Pt break-junction experiments \cite{Thijssen2006}, with a different fluctuation behavior of hydrogen and deuterium. We further note that perp I and II converge to the same structure at close contact ($G \approx 0.6$~G$_0$, Fig.~\ref{theory_sup5} in Appendix A).

The importance of the NQE is substantiated by the significant vibrational quantum zero-point energy (ZPE) calculated in the harmonic approximation, which is the same order of magnitude as the energy difference between the FeH$_2$ perp I and para configurations ($\sim 0.1$ eV, see Table~\ref{table1} and Fig.~\ref{theory_sup2} in Appendix B). 
The FeH$_2$ para has a lower total energy for frozen ions, but its energy is approximately 0.08~eV higher than the perp I configuration when including the ZPE.
 
\begin{figure} [h]
\includegraphics[width=0.9\columnwidth]{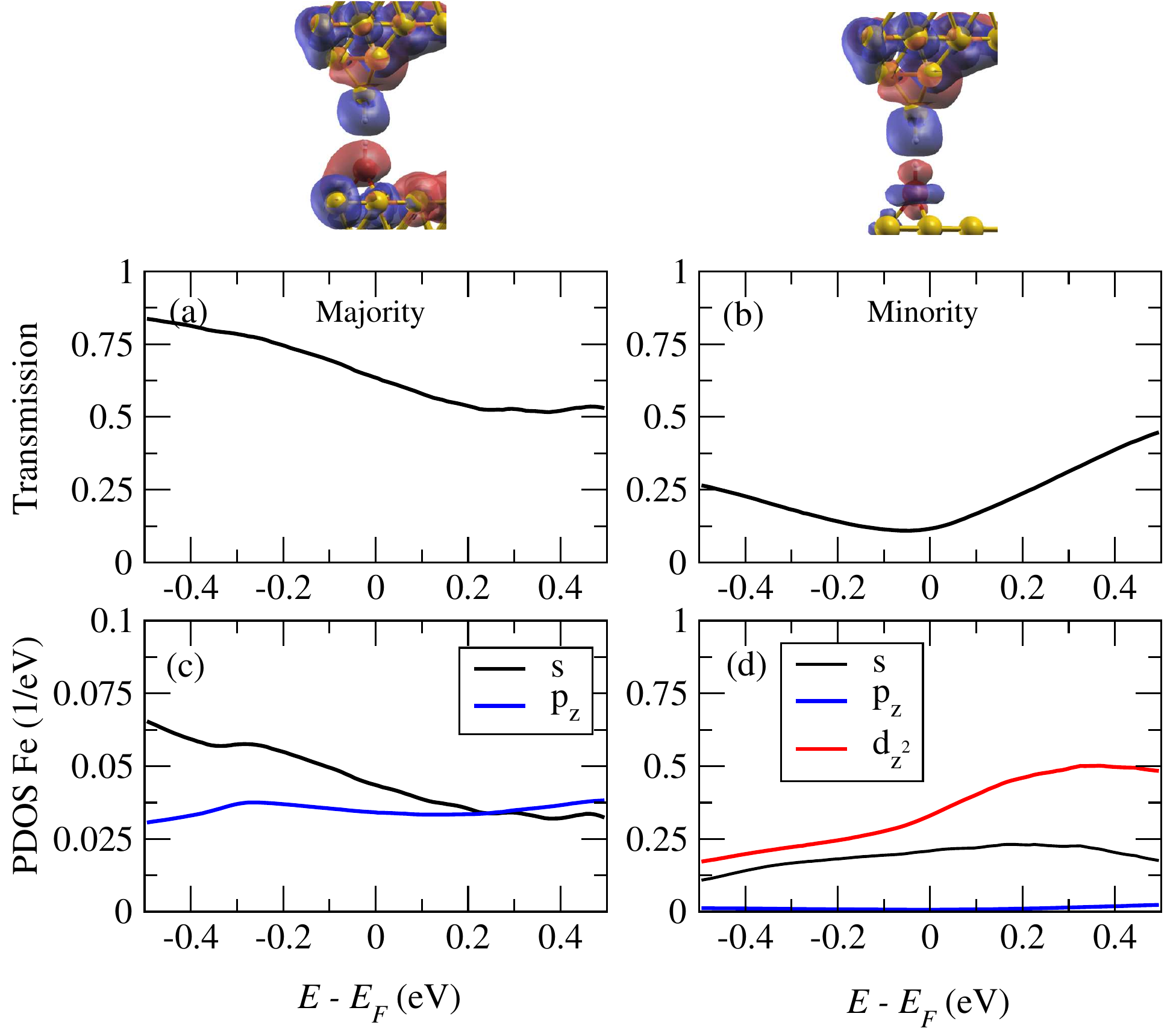}
\caption{Eigenchannel wave functions (real part, incident from the tip) for majority (left) and minority (right) channels. 
(a) Majority and (b) minority spin transmissions (single channel) for the FeH$_2$ perp I configuration along with (c, d) the corresponding projected DOS on Fe. 
The $z$-axis is orientated along the contact.}
\label{theory_fig3}
\end{figure}

We can gain more insights when comparing the calculated Fano factor to the experiments in Fig.~\ref{theory_fig2}(b). Here we observe that for FeH$_1$, the $P$ becomes almost 100\% in the low conductance regime ($G < 0.2$~G$_0$), while it yields a high Fano factor after the conductance plateau ($G > 0.5$~G$_0$) at variance with the experiments.
On the other hand the FeH$_2$ perp I gives the best agreement with experimental observations, namely, a clear enhancement of the $P$ up to about 80\% in the case of hydrogenated Fe contacts, while the FeH$_2$ perp II yields nearly the same $P$ as for clean Fe. 
This is attributed to multichannel transport behavior in the contact regime, as shown in Fig.~\ref{theory_sup7} of Appendix C.  For the parallel FeH$_2$ configurations (here, two possible configurations are considered), we found almost zero spin-polarized transport (Appendix A, Fig.\ref{theory_sup3}) and, combined with the LDOS-height, these configurations seems not to fit the experimental data.

The reason for the high transport spin polarization for the FeH$_2$ perp I configuration is twofold. Firstly, the hydrogen molecule in the perpendicular configuration with its rotational symmetry around the $z$-axis only allows transmission of rotationally symmetric states, namely the $s$, $d_{z^2}$, and $p_z$ orbitals on the Fe. This is reflected in the eigenchannel wave functions in Fig.~\ref{theory_fig3}, which show that the transport takes place mainly through the H$_2$-LUMO orbital and that the Fe $s$-orbital is involved for the majority channel, while also the Fe $d_{z^2}$-orbital is involved for the minority channel. Secondly, the minority transmission dip is due to the destructive quantum interference between the pathways from H$_2$-LUMO via either $s$ and $d_{z^2}$ states, as explained with a toy model in Fig.~\ref{theory_sup9} of Appendix C.
 
\section{VI. Discussion}

Hydrogen and deuterium have been studied in a number of metal break-junction experiments along with vibrational spectroscopy and noise measurements \cite{Li2015, Djukic2006, Kiguchi2010, Smit2002, Thijssen2006}. 
Hydrogen molecules in Pt junctions have previously been shown to lead to conductances close to the conductance quantum G$_0$ \cite{kig07}.
For the H$_2$-Pt the perpendicular structure has been advocated by the noise measurements indicating a single channel in agreement with DFT calculations \cite{Kristensen2009}.
Other DFT calculations, using `cluster-electrodes', yield rather similar energies for the two configurations \cite{Garcia2004}.
We note that although first principles calculation have reproduced the single-channel transmission and behavior of the vibrational frequencies with stretching of the Pt contact, the sign of the inelastic signals disagree with the experiments \cite{Kristensen2009}.
The cause for this may be the missing NQE and fluctuations mentioned above.
Moreover, on Au(110) the hydrogen vibrational energies have been observed to vary as the tip of a STM was brought closer to the molecule \cite{Hauho15}.
DFT calculations suggested that the tip initially attracts the molecule, which lies flat on the surface, and tends to turn it upright.
At further reduced tip-surface separations, H$_2$ was displaced in a lateral direction in the calculations.

In the present case we consider spin-polarized systems where one electrode is well-characterized prior and after the contact formation. 
Our experiments and calculations consistently show a modest spin polarization of the Fe contacts.
In the experiments, hydrogen drastically increases the apparent height of the Fe adatom.
As for the Fano factor, we find some reduction compared to pristine Fe, in particular at low conductances.

In the calculations, some properties are hardly affected by hydrogen.
The magnetic moment of the Fe adatom varies between 3.2 -- 3.7~$\mu_B$ depending on the contact geometry.
Moreover, the conductance of the second plateau at close separations remains close to $\approx 0.7$~G$_0$, the value observed from Fe.
Both H and H$_2$ perp II  introduce a conductance plateau near $\approx 0.3$~G$_0$. 

Only a vertical H$_2$ molecule reproduces the experimental height change. 
In this orientation, the molecule reduces the calculated Fano factor at small conductances, in agreement with the experimental result.
This reduction is due to orbital symmetries and an interference effect. 
Experimental indication of the importance of orbital symmetry for spin polarization has been reported in Ref.~\onlinecite{vardimon2015indication} based on conductance data from NiO contacts.
Such an effect has also been found in transport calculations \cite{jac6, roc7}. 
Quantum interference has been invoked to interpret the high $P$ from benzene and vanadocene molecules in break junctions \cite{Dongzhe2019, pal19}.
The present case is remarkable because the addition of a simple hydrogen molecule combines both effects.
	
The influence of a parallel hydrogen molecule is strikingly different.
The Fano factor is increased compared to Fe at all conductances.
At low conductances, the Fano factor evolves on the $F(G)$ line expected for unpolarized transport through a single channel.
At more elevated $G$ ($\approx 0.5$~G$_0$ for para I, $\approx 0.6$~G$_0$ for para II), the Fano factor exceeds this line, which is clear evidence of transport through several channels.
 
\section{VII. Summary}
 
Pristine and hydrogenated Fe adatoms on Au(111) have been imaged with low temperature STM\@.
Contacts to these atoms and molecules have been reproducibly made and the shot noise of the current has been measured.
DFT-based transport calculations reproduce several aspects of the experimental data.
In particular, the modeling suggests that the hydrogen molecule in the STM contact is preferentially oriented perpendicular to the surface.
For conductances below a conductance quantum we find a single relevant conductance channel that enables spin-polarized transmission through Fe atoms. 
Hydrogen increases the spin polarization up to 80\% via the effects of orbital symmetries and interference. 
Despite the partial agreement of the experimental and theoretical data, some deviations remain. 
The quantum motion of hydrogen atoms, which is beyond our DFT modeling, is a likely reason for the discrepancies.
 
\section{Acknowledgements}
 
We thank the Deutsche Forschungsgemeinschaft for support via project BE2132/8-1. This project received funding from the EU Horizon 2020 under Grant No.~766726. Part of the calculations were done using HPC resources from CALMIP (Grant No.~P21023).
\appendix
\section{APPENDIX A: FANO-FACTOR AND CONDUCTANCE ANALYSIS FROM DFT+NEGF}
The total energy of both perpendicular FeH$_2$ structures is shown in Fig. 10 for different tip-adatom distances. The calculated conductance of FeH$_1$ is plotted in Fig. 11 versus tip apex to adatom distance. Here a conductance plateau and the yielding (relaxation) of the Fe-substrate bonds can be observed (compare with FeH$_1$ curve in Fig. 7(a)).

\begin{figure}[h]
	\centering
	\includegraphics[width=.9\columnwidth]{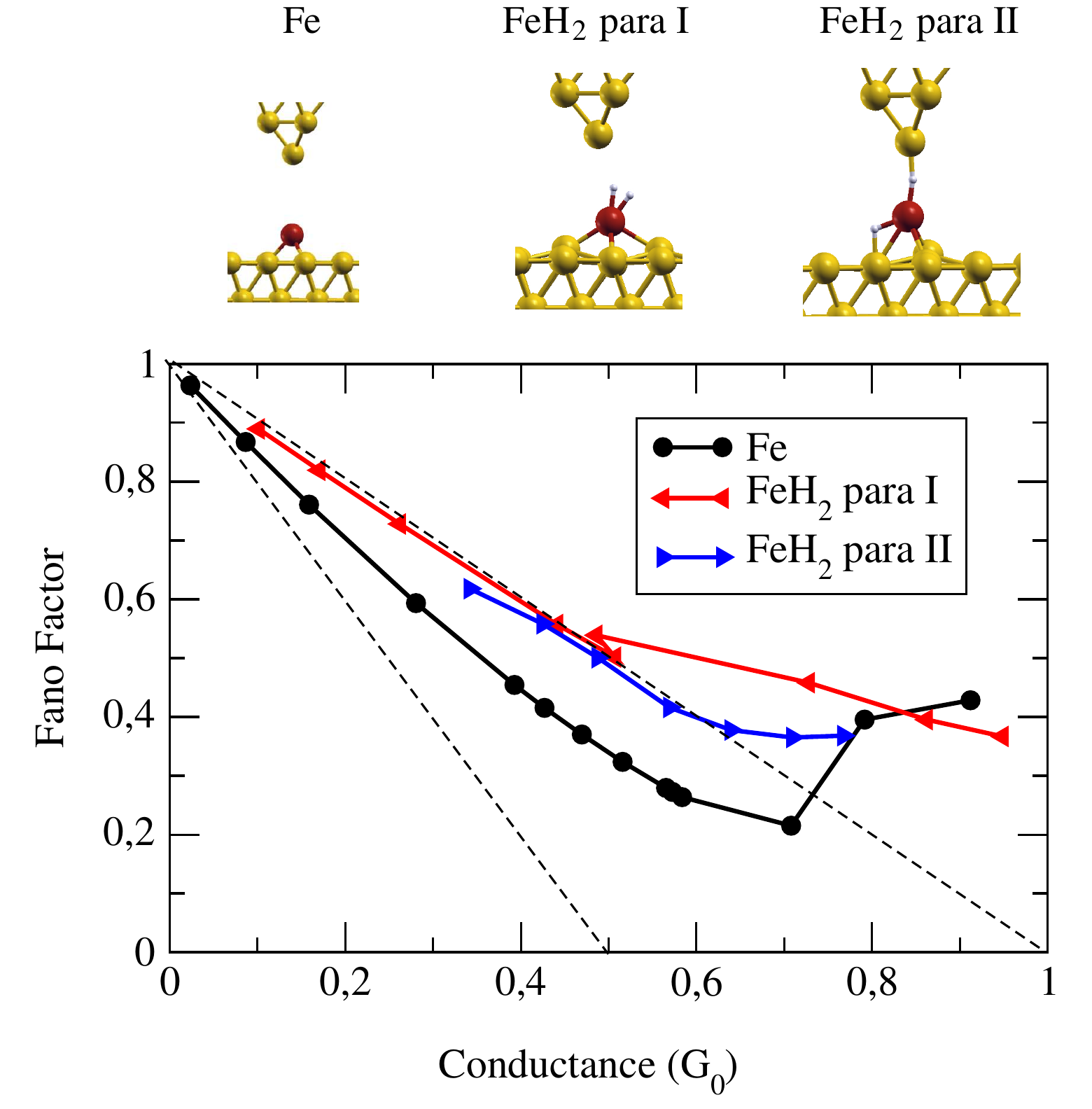}
	\caption{
		Calculated Fano factors versus conductance for Fe (black circles), FeH$_2$ para I (red left triangles) and FeH$_2$ para II (blue right triangles) configurations.}
	\label{theory_sup3}
\end{figure}\textbf{}

\begin{figure}[h]
	\centering
	\includegraphics[width=0.9\columnwidth]{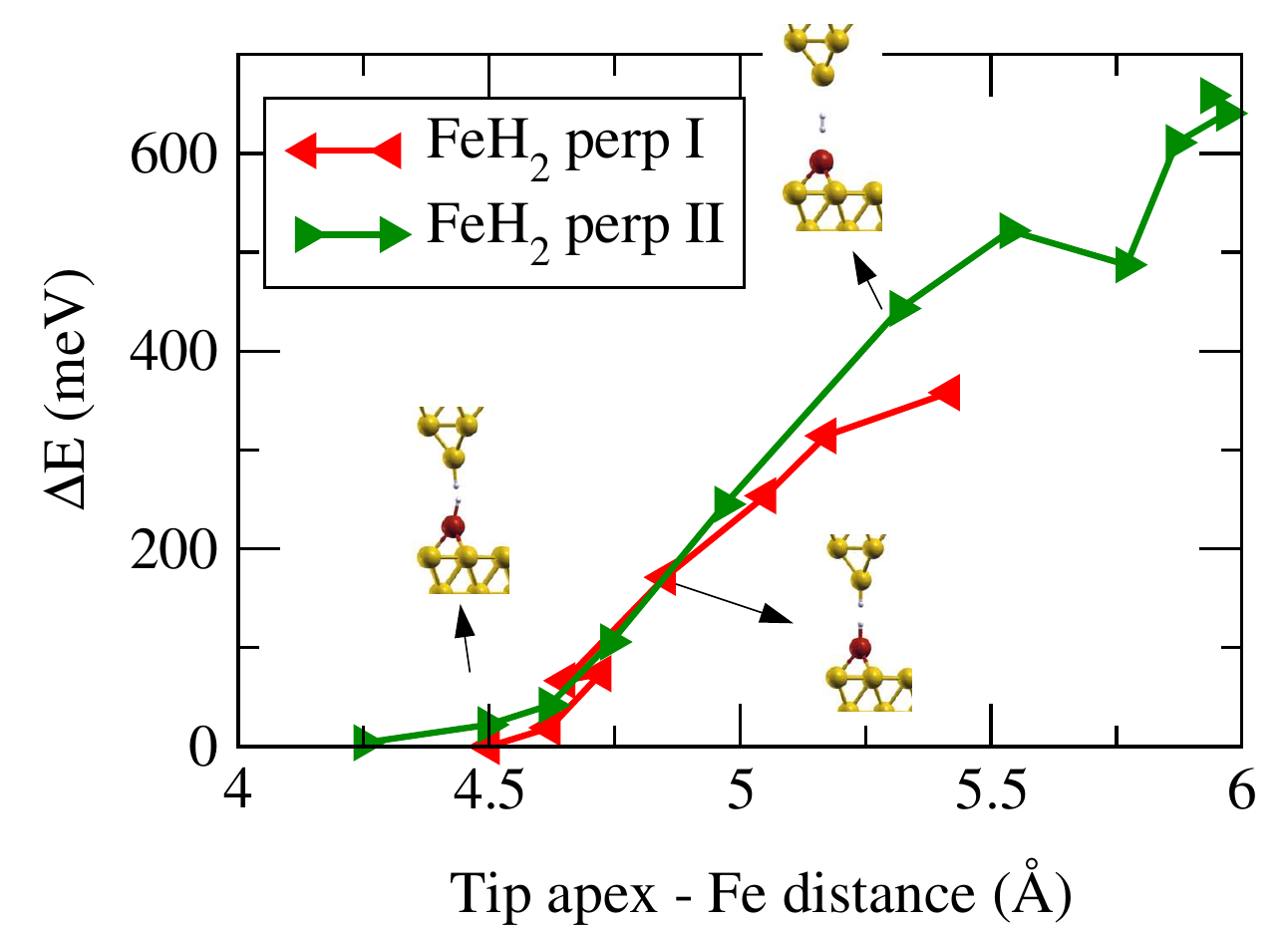}
	\caption{
		Total energy variation with respect to the tip apex - Fe distance. The leftmost point of FeH$_2$ perp I was set as zero.}
	\label{theory_sup5}
\end{figure}

\begin{figure}[h]
	\centering
	\includegraphics[width=\columnwidth]{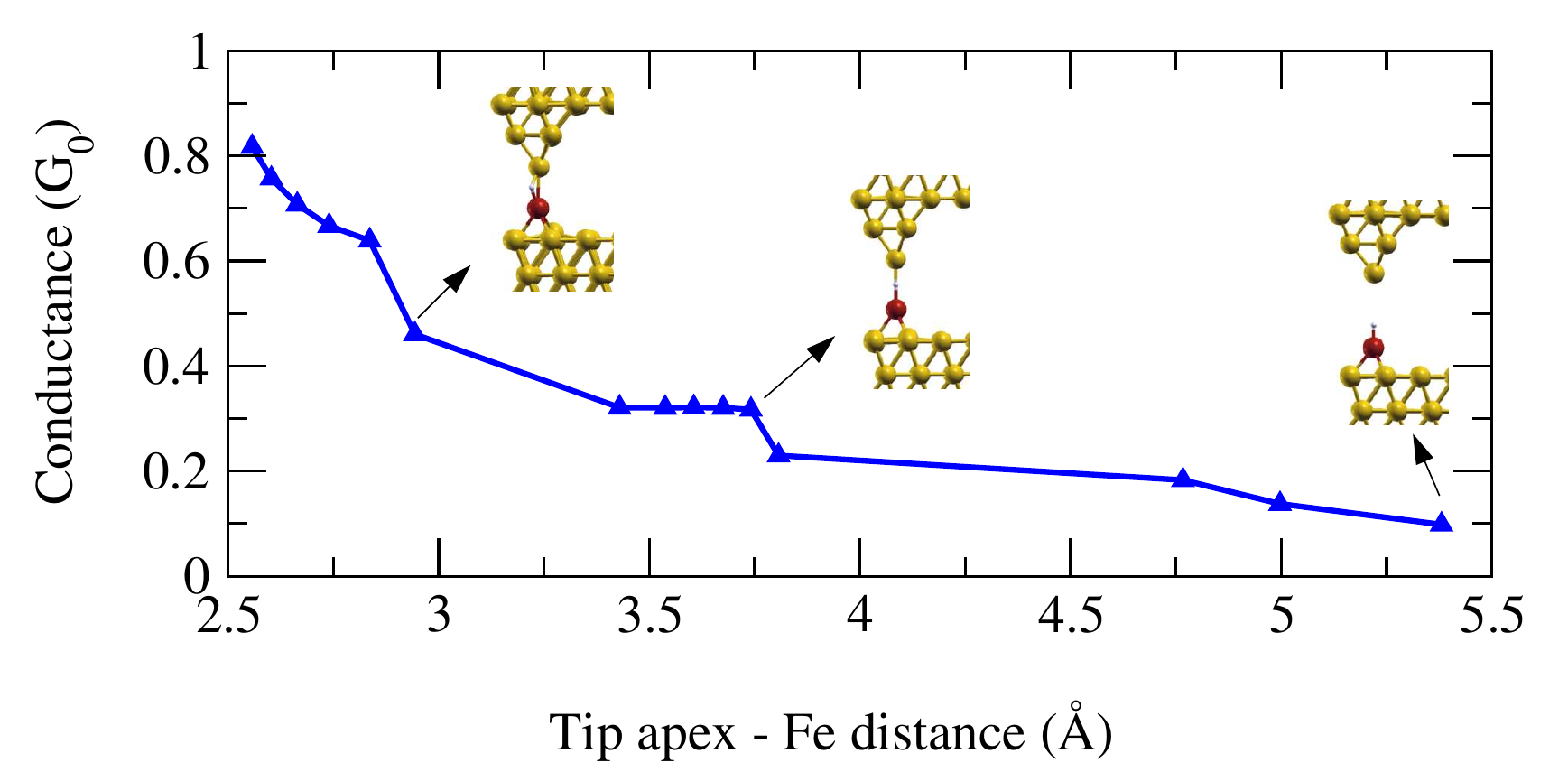}
	\caption{Conductance variation during the approach process for FeH$_1$ contacts. A clear conductance plateau is observed.}
	\label{theory_sup11}
\end{figure}

\section{APPENDIX B: HYDROGEN VIBRATIONS}
Figure 12 shows the caclculated phonon modes for FeH$_2$ perp I and FeH$_2$ para I, whose energies are listed in Table I. 
\begin{figure}[h]
\centering
	\includegraphics[width=1\columnwidth]{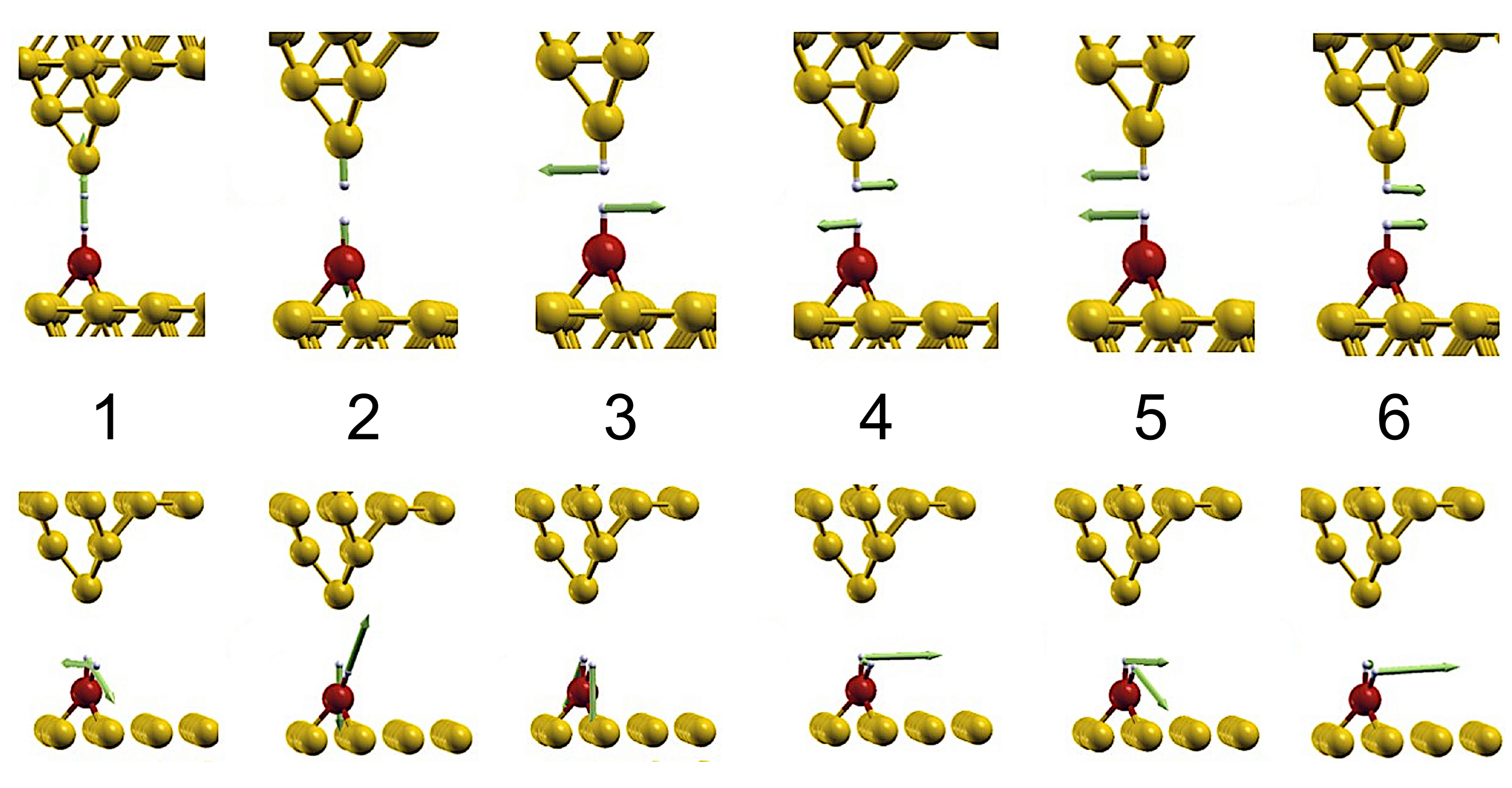}
	\caption{	Calculated phonon modes for FeH$_2$ perp I (top) and para I (bottom).
The corresponding zero-point energies are $E_{zp}=(1/2)\sum_n \hbar\omega_n$ and 260 meV and 364 meV, respectively.
}
\label{theory_sup2}
\end{figure}
\FloatBarrier

\begin{table}[h]
	\centering
	\scalebox{1}{
		\begin{tabular}{ccccc}
			\hline\hline
			\multicolumn{1}{c}{$$} &\multicolumn{1}{c}{$$} & \multicolumn{1}{c} {FeH$_2$ perp I (meV)} &\multicolumn{1}{c}{$$}& \multicolumn{1}{c} {FeH$_2$ para I (meV)}   \\ 
			1  &  & 182.78 && 304.82  \\
			2  &  & 98.92 && 193.67  \\
			3  &  & 79.21 && 124.01    \\
			4 &  & 79.00  && 44.87    \\
			5 &  & 41.09 && 39.17  \\
			6  &  &40.90 & &21.30  \\ \hline
	\end{tabular}}
	\caption{Calculated phonon modes for FeH$_2$.} 
	\label{table1}
\end{table}	

\section{APPENDIX C: TRANSMISSION ANALYSIS AND TIGHT-BINDING TOY MODEL}	
The transmission functions of the relevant transport eigenchannels of Fe, FeH$_1$, FeH$_2$ perp I, and FeH$_2$ perp II are displayed in Fig. 13. Destructive quantum interference between transport pathways via Fe $s$ and $d_{z^2}$ states reduce the transmission as reproduced within a tight-binding description shown in Fig. 14.
\begin{figure}[!h]
	\centering
	\includegraphics[width=1\columnwidth]{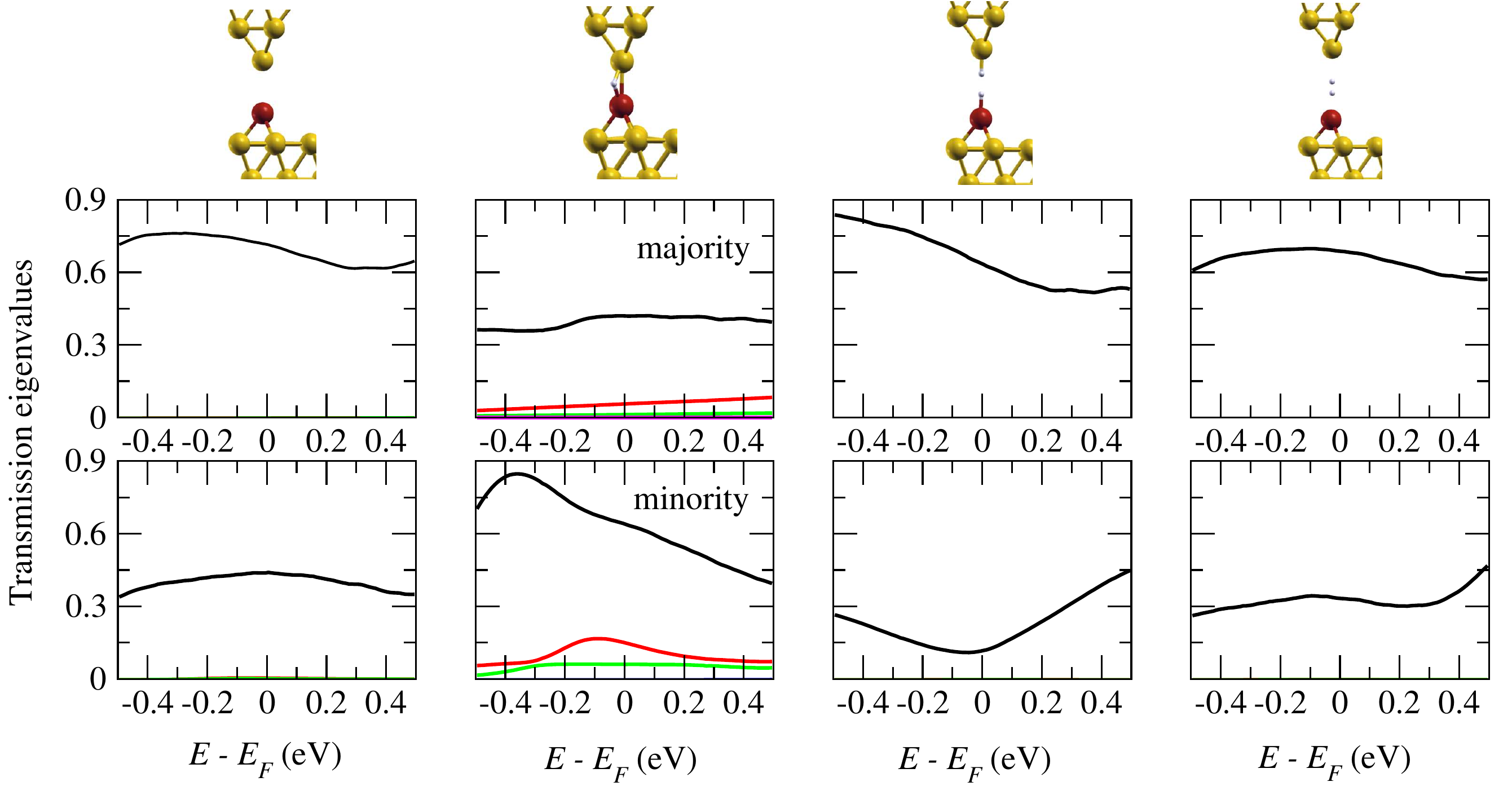}
	\caption{Transmission eigenchannels for Fe, FeH$_1$, FeH$_2$ perp I, and FeH$_2$ perp II close to contact. Colors indicate different transport channels. Only FeH$_1$ shows multichannel transport behavior.}
	\label{theory_sup7}
\end{figure}

\begin{figure} [h]
    \centering
    \includegraphics[width=0.75\columnwidth]{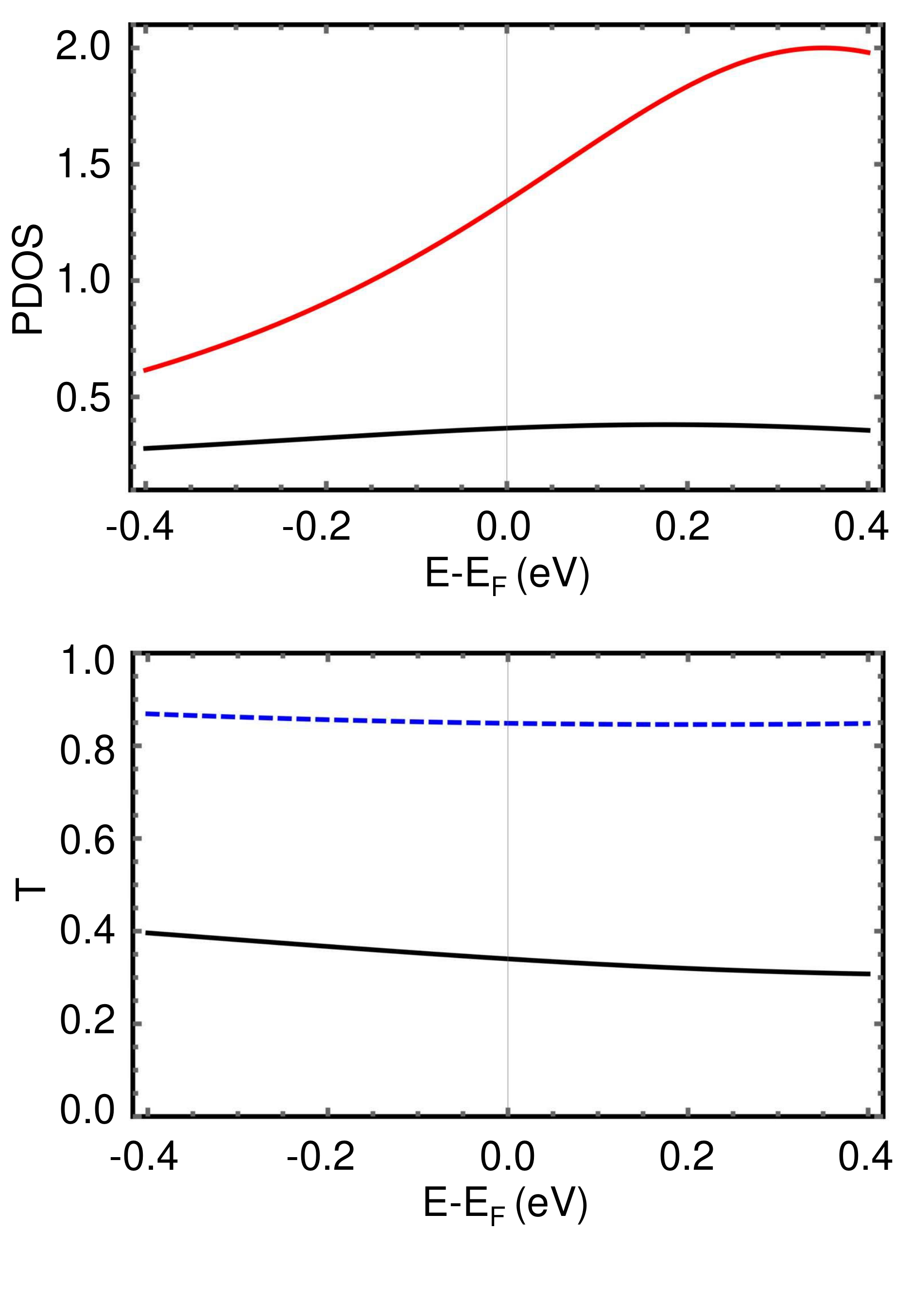}
    \caption{Simplified three level model of interference between two paths from H$_2$-LUMO via Fe $s$ or $d_{z^2}$ to the surface. Upper panel: PDOS of Fe $s$ (black) and Fe $d_{z^2}$ (red) orbitals, respectively. Lower panel: Transmission without coupling to the $d_{z^2}$-level (blue dashed) and transmission where both paths are included (black). Parameters (eV): Fe coupling to surface: $\Gamma_s=3$, $\Gamma_d = 1.0$, Fe levels: $\varepsilon_s=-0.3$, $\varepsilon_d = 0.35$, H$_2$-LUMO coupling to tip $\Gamma_t=4$, level $\varepsilon_{\text{LUMO}}=0.5$. Coupling of LUMO to $s$, $d$ respectively, $t_s=3$, $t_d=1.8$ (black) $t_d=0$ (blue dashed).}
    \label{theory_sup9}
\end{figure}

\FloatBarrier

\end{document}